\documentclass[12pt]{article}
\usepackage{setspace}
\usepackage{natbib}
\usepackage{graphicx}
\usepackage{lscape}
\usepackage{pdflscape}
\usepackage{afterpage}
\usepackage{pdfpages}
\usepackage{float}
\usepackage{booktabs}
\usepackage[utf8]{inputenc}
\usepackage{indentfirst}
\usepackage{arydshln}
\usepackage{arydshln}
\usepackage{tikz}
\usepackage{lipsum}
\usepackage{pgffor}
\usepackage{dsfont}
\usepackage{amsfonts}
\usepackage{amsmath}
\usepackage{xfrac}
\usepackage{amssymb}
\usepackage{graphicx}
\usepackage{url}				
\usepackage{subfig} 
\usepackage{thmtools}
\usepackage{soul}

\renewcommand\thmcontinues[1]{continued}

\usepackage{graphicx}

\newtheorem{theorem}{Theorem}

\newtheorem{corollary}{Corollary} 

\newtheorem{example}[theorem]{Example}

\newtheorem{proposition}{Proposition} 
\newtheorem{remark}{Remark} 

\newenvironment{proof}[1][Proof]{\noindent \textbf{#1.} }{\  \rule{0.5em}{0.5em}}

\newtheorem{assumption}{Assumption} 

\usepackage{hyperref} 
\hypersetup{
	colorlinks=true, breaklinks=true, bookmarks=true,bookmarksnumbered,
	urlcolor=red, linkcolor=red, citecolor=blue, 
	pdftitle={}, 
	pdfauthor={\textcopyright}, 
	pdfsubject={}, 
	pdfkeywords={}, 
	pdfcreator={pdfLaTeX}, 
	pdfproducer={LaTeX with hyperref and ClassicThesis} 
}

\usepackage[top=1in, bottom=1in, left=1in, right=1in]{geometry}

\usepackage{bbm}

\begin{document}

	\def\spacingset#1{\renewcommand{\baselinestretch}%
		{#1}\small\normalsize} \spacingset{1}

	
\title{ { \LARGE Extensions for Inference in Difference-in-Differences with Few Treated Clusters }}

\author{
Luis Alvarez\footnote{Sao Paulo School of Economics - FGV; email: luisfantozzialvarez@gmail.com } \and Bruno Ferman\footnote{Sao Paulo School of Economics - FGV; email: bruno.ferman@fgv.br} 
}

\date{}
\maketitle

	\newsavebox{\tablebox} \newlength{\tableboxwidth}
	

	\begin{center}


\

\

\

\

\textbf{Abstract}

\end{center}

In settings with few treated units, Difference-in-Differences (DID) estimators are not consistent, and are not generally asymptotically normal. This poses relevant challenges for inference. While there are inference methods that are valid in these settings, some of these alternatives are not readily available when there is variation in treatment timing and heterogeneous treatment effects; or for deriving uniform confidence bands for event-study plots. We present alternatives in settings with few treated units that are valid with variation in treatment timing and/or that allow for uniform confidence bands.

\

	\noindent%

	\
	
	\noindent%
		
		\vfill

	\newpage
	\spacingset{1.45} 
	

\onehalfspacing

\section{Introduction}

In settings with few treated units, Difference-in-Differences (DID) estimators are not consistent, and are not generally asymptotically normal, posing relevant challenges for inference \citep{Donald,CT}. In such settings, relying on standard procedures, such as clustering the standard errors at the unit level, may lead to severe over-rejection. In some examples, we may expect rejection rates on the order of more than 60\% for a 5\% nominal-level test \citep{Assessment}. In light of these concerns, some alternative inference methods have been proposed  for settings in which we have a small number of treated units \citep{CT,FP,MW,hagemann2020inference}. 

However, some of these alternatives are not readily available to incorporate some recent advances in the DID literature. Such advances include (i) considering settings with heterogeneous treatment effects and  variation in the treatment timing \citep{Bacon,clement2,Chaisemartin,Pedro,SUN2020}, and (ii) relying on uniform confidence bands when presenting event-study plots \citep{Freyal2021,Pedro}.

In this note, we extend the inference methods proposed by \cite{CT} and by \cite{FP} so that (i) we can incorporate the recent recommendations from papers that analyzed settings with heterogeneous treatment effects and  variation in the treatment timing, and/or (ii) we can allow for uniform confidence bands when considering event-study plots.  \cite{CT} discuss in their Section III.A the possibility of extending their approach to heterogeneous treatment effects. In this note, we formalize this idea by filling in important implementation details, and we  show how parametric models for heteroskedasticity, as proposed by \cite{FP},  can be extended to a staggered adoption setting. \cite{CT} also discuss the possibility  of  computing joint confidence sets based on the inversion of a test statistic. Differently, we show how one can conduct uniform inference in this setting by relying on uniform confidence bands, which have been advocated in the literature due their ease of interpretability and computation \citep{Montiel2018}. Importantly, due to the nonstandard nature of the setting, critical values used in the uniform bands are not based on a known distribution. We show that a specific bootstrap algorithm is able to recover the required critical values in an asymptotic framework where the number of treated units is fixed and the number of controls is large.


\section{Setting}

 Let $y_{j,t}(0)$ ($y_{j,t}(1)$) be the potential outcome of unit $j$ at time $t$ when this unit is untreated (treated) at this period. We consider first that potential outcomes are given by
\begin{eqnarray} \label{simple_did_equation}
\begin{cases} y_{j,t}(0) =   \theta_j + \gamma_t + \eta_{j,t}  \\  y_{j,t}(1) = \alpha_{j,t} +y_{j,t}(0),  \end{cases} 
\end{eqnarray}
where $\theta_j$ and $\gamma_t$ are, respectively, unit- and time-invariant unobserved variables, while $\eta_{j,t}$ represents unobserved variables that may vary at both dimensions; $\alpha_{j,t}$ is the (possibly heterogeneous) treatment effect on unit $j$ at time $t$. We consider $ \alpha_{j,t}$ as a fixed parameter, which means that we define the target parameters based on the realized treatment effect of this policy for the treated units. In this case, uncertainty regarding this parameter comes from unobserved shocks that may affect the potential outcomes of unit $j$ at time $t$, such as, for example, weather or economic shocks that are unobserved by the econometrician ($\eta_{j,t}$).  A setting in which treatment assignment and treatment effects are treated as fixed is common in the literature of DID with few treated clusters \citep{CT,FP,ferman2020inference}. A similar setting is also considered in other settings in which the number of treated clusters is fixed, such as in the synthetic controls literature \citep{Abadie2010,SDID, ASC,CWZ,FP_QE,Ferman_JASA}.

Units $j = 1,...,N_1$  are treated at some point, while units $ j = N_1 + 1,...,N$ are never treated, where $N_0 = N-N_1$. We observe information for periods $t=1,...,T$, and we allow for variation in treatment timing by denoting $t^\ast_j \in \{1,...,T-1\}$ as the last period before unit $j$ enters into treatment, for $j \leq N_1$. Let $t^\ast_j = \infty$ for $j > N_1$.  Treatment is assumed to be an absorbing state and we assume there is no anticipation, so observed outcomes are given by $y_{jt} = \mathbf{1}\{t>t_j^*\}y_{jt}(1) + \mathbf{1}\{t\leq t_j^*\}y_{jt}(0) $.

We define the target parameter $\bar \alpha = \sum_{i = 1}^{N_1} \sum_{t = t_{j}^* + 1}^T \omega_{j,t} \alpha_{j,t}$, which is a  linear combination of the treatment effects of different units in different periods. One example for a target parameter is the average treatment effects for the treated units in the periods that they were treated. Alternatively, we can consider, for example, weighted averages depending on populations sizes. 

We might also be interested in a multivariate parameter $\bar{\boldsymbol{\alpha}} = (\bar \alpha_1, ..., \bar \alpha_K)$, in which case we set $\bar \alpha_k = \sum_{i = 1}^{N_1} \sum_{t = t_j^*+1}^T \omega_{k,j,t} \alpha_{j,t}$. For example, this may include the treatment effects $\tau$ periods after the start of the treatment. With some abuse of notation, we can also consider that some of these $\bar \alpha_k$ include pre-treatment trends, which are commonly presented in dynamic DID models as an assessment for the parallel trends assumption. We provide an example of the latter in the next sections.

Let $\boldsymbol{\eta}_j =( \eta_{j,1},...,\eta_{j_T})'$. Uncertainty comes from different realizations of $\{\boldsymbol{\eta}_j \}_{j=1}^N$, where we treat treatment allocation as fixed. We also consider the possibility of a vector of observable variables $Z_j$ that may be determinants of the heteroskedasticity, as we discuss below. These variables are also treated as nonrandom throughout.\footnote{We can also consider the case in which some of these variables enter in the model for $y_{it}(0)$ in \eqref{simple_did_equation}. }

\begin{assumption}{(Sampling)}
\label{Assumption_sampling}
For each $j = 1,...,N$, we observe $(y_{j,1},...,y_{j,T},Z_j)$, where $y_{j,t} = y_{j,t}(0) $ if $t < t^\ast_j $, and $y_{j,t} = y_{j,t}(1)$ otherwise. Potential outcomes are determined by Equation  \ref{simple_did_equation}. We assume that $(\boldsymbol{\eta}_1,...,\boldsymbol{\eta}_N)$ are mutually independent, with finite second moments.

\end{assumption}

We do not need to impose any assumptions on $\theta_j$ and $\gamma_t$.

\section{Estimator}

 An important limitation of the TWFE estimator in this setting is that it may recover a linear combination of the treatment effects $\alpha_{j,t}$ in which some of the weights might be negative \citep{Chaisemartin,Bacon,BJS}. Most of the solutions in this case consider alternative estimators that combine simpler $2 \times 2$ estimators \citep{SUN2020,Pedro,BJS}.\footnote{Indeed, \cite{BJS} show that, under Assumptions \ref{Assumption_sampling} and \ref{Assumption_PT}, all linear-in-outcomes unbiased estimators of $\boldsymbol{\bar{\alpha}}$ take an ``imputation'' form, of which aggregation of simple 2 $\times$ 2 DID estimators constitute a particular case.}$^,$\footnote{\cite{Pedro} consider a doubly robust estimation method for each of those $2 \times 2$ estimators, instead of considering simple DID estimators. Even though their estimator is not consistent in our setting, our inference method remains valid for weighted averages of the treatment effect, where the weights are given by the estimated propensity score, provided that the outcome model is correctly specified. Indeed, their estimator is not doubly robust in our setting with a few number of treated units.}   We follow a similar approach. More specifically, for each $\alpha_{j,t}$ in which $\omega_{k,j,t}>0$ for some $(k,j,t)$, we consider an estimator $\hat \alpha_{j,t}$, and then we aggregate them to get $\widehat{\bar \alpha}_k =  \sum_{i = 1}^{N_1} \sum_{t = t_j^* +1}^T \omega_{k,j,t} \hat \alpha_{j,t}$. We note that \cite{CT} already recommended estimating each $\alpha_{j,t}$ separately in settings with heterogeneous treatment effects, even before the recent papers that pointed out the aggregation problems of the TWFE estimator.  

For a given $(\nu_{1}(j,t),...,\nu_{t^\ast_j}(j,t))$ that satisfies $\sum_{\tau=1}^{t^\ast_j} \nu_{\tau}(j,t) = 1$, we consider estimators  $\hat \alpha_{j,t}$ of the form
\begin{eqnarray} \label{generic}
\hat \alpha_{j,t} = \left[ y_{  j,t} - \sum_{\tau=1}^{t^\ast_j} \nu_\tau(j,t) y_{  j,\tau}\right] - \frac{1}{N_0} \sum_{j' = N_1 + 1}^N \left[ y_{  j',t} - \sum_{\tau=1}^{t^\ast_j} \nu_\tau(j,t)  y_{  j',\tau}  \right].
\end{eqnarray}

That is, for the post-pre comparison, we compare period $t$ with a weighted average of the pre-treatment periods given by the weights $\nu_\tau(j,t)$. Then we compare this post-pre comparison for the treated unit $j$ with the average of the controls.\footnote{We do not consider the possibility of using the not-yet-treated as controls, because this would  be asymptotically irrelevant when $N_1$ is fixed and $N_0 \rightarrow \infty$, and because this would make the notation more complicated. Still, it is possible to consider this alternative. } Two simple examples include
\begin{eqnarray}
\hat \alpha_{j,t} = \left[ y_{  j,t} - \frac{1}{t^\ast_j }\sum_{\tau=1}^{t^\ast_j} y_{  j,\tau}\right] - \frac{1}{N_0} \sum_{j' = N_1 + 1}^N \left[ y_{  j',t} - \frac{1}{t^\ast_j }\sum_{\tau=1}^{t^\ast_j} y_{  j',\tau}  \right],
\end{eqnarray}
where we consider a DID estimator using unit $j$ and the never treated, for all pre-treatment periods and for period $t$. Alternatively, we can use only the last period  before unit $j$ starts  treatment as the pre-period, so that 
\begin{eqnarray}
\hat \alpha_{j,t}' = \left[ y_{  j,t} - y_{  i,t^\ast_j}\right] - \frac{1}{N_0} \sum_{j' = N_1 + 1}^N \left[ y_{  j',t} - y_{ j',t^\ast_j}  \right].
\end{eqnarray}

We aggregate these estimators the following way. Let $\mathbf{Y}_j = (y_{j,1},...,y_{j,T})$. For each $j =1,\ldots, N_1$, we define a $(K_j \times T)$ matrix $A_j$ such that $A_j[ \mathbf{Y}_j - \frac{1}{N_0}\sum_{i=N_1+1}^{N} \mathbf{Y}_i]$ consists of stacked estimators of $\hat{\alpha}_{jt}$ for a subset of the periods. We provide examples of choices of $A_j$ below. We then aggregate these building-block estimators onto a $K$-dimensional estimator of $\bar{\boldsymbol{\alpha}} $ through $(K \times K_j)$ matrices $B_j$. The resulting estimator is given by $\widehat{\bar{\boldsymbol{\alpha}}}\equiv \sum_{j=1}^{N_1} B_j A_j \left[\mathbf{Y}_j  - \frac{1}{N_0}\sum_{i=1}^{N_0}\boldsymbol{Y}_i\right] $, whereas the target parameter may be written as $\bar{\boldsymbol{\alpha}} = \sum_{j=1}^{N_1} B_j A_j \boldsymbol{\alpha}_j$, where $\boldsymbol{\alpha}_j = (0,\ldots,0, \alpha_{j,t^*_j+1},\ldots, \alpha_{jT})'$ is the vector of identifiable treatment effects for unit $j$.

 We consider the following parallel trends assumption.

\begin{assumption}{(Parallel trends)}
\label{Assumption_PT} For each $j =1,\ldots, N_1$, we assume that:

$$B_j A_j \mathbb{E}[\boldsymbol{\eta}_{j}]  = B_j A_j \mathbb{E}[\boldsymbol{\eta}_{i}], \ \forall i \in \{N_1+1,\ldots, N \}\, .$$

We also assume that each row of $B_j A_j$ sums up to zero.
\end{assumption}

Assumption \ref{Assumption_PT} guarantees the relevant parallel trends conditions for the estimator $\widehat{\bar \alpha}$ of our choice. If we use all pre-treatment periods the estimation of each $\alpha_{j,t}$, then Assumption \ref{Assumption_PT} in practice means that we have parallel trends for all periods.\footnote{Assumption \ref{Assumption_PT} would actually be weaker than that, as it may be satisfied without assuming parallel trends for all periods. In this case, however, we would need  departures from parallel trends to cancel out, so that this assumption is satisfied. } In contrast, if we consider the estimator $\hat \alpha_{j,t}'$, then we only need parallel trends between period $t$ and the last period before treatment for unit $j$. The requirement that the rows of $B_j A_j$ sum up to zero is made so the estimator removes unit fixed effects $\theta_j$, and it is satisfied for our generic estimator presented in Equation \ref{generic}, given the restriction $\sum_{\tau=1}^{t^\ast_j} \nu_{\tau}(j,t) = 1$.

\begin{example}[label=exa:scalar]
	\label{example_scalar}
	Suppose we are interested in the average post-treatment effect on the treated units, and that we use all pre-treatment periods in constructing our estimator.  For $j\leq N_1$, put $K_j = T - t_j^*$. In this case, the matrices $A_j$ are taken as
	
	\begin{equation}
		A_j = \begin{bmatrix}
			- \frac{1}{t_j^*}   \mathbf{1}_{K_j \times t_{j}^* } & \mathbb{I}_{K_j}
		\end{bmatrix},
	\end{equation}
where $\mathbf{1}_{a \times b}$ is a $a \times b$ matrix of ones and $\mathbb{I}_{c}$ is a $c \times c$ identity matrix. The matrices $B_j$ are taken as

$$B_j = \frac{1}{\sum_{i=1}^{N_1}K_i} \mathbf{1}_{1 \times K_j}\, ,$$
and Assumption \ref{Assumption_PT} requires that, for each $j = 1,\ldots, N_1$,

$$\sum_{t=t_{j}^*+1}^T \frac{\mathbb{E}[\eta_{jt}]}{(T- t_j^*)} -  \sum_{t=1}^{t_j^*} \frac{\mathbb{E}[\eta_{jt}]}{t_j^*} = \sum_{t=t_{j}^*+1}^T \frac{\mathbb{E}[\eta_{it}]}{(T- t_j^*)} -  \sum_{t=1}^{t_j^*} \frac{\mathbb{E}[\eta_{it}]}{t_{j}^*}, \forall i \in \{N_1+1,\ldots N_0\}\, .$$
\end{example}

\begin{example}[label=exa:multiv]
	\label{example_es}
	Suppose that we are interested on average treatment effects by length of exposure. Suppose that we use all pre-treatment periods in constructing our estimator. In this case, the $A_j$ can be taken as in the previous example. The $B_j$ are constructed as follows. For each $k \in \mathbb{N}$, define $N_k = \sum_{i=1}^{N_1} \mathbf{1}_{k \leq T - t_{j}^*}$. Let $K = \max_{i=1,\ldots,N_1} K_i$. We define $B_j$ as:
	
	\begin{equation}
		B_j = \begin{bmatrix}
			\operatorname{diag}(1/N_1, 1/N_2, \ldots, 1/{N_{K_j}}) \\
			\boldsymbol{0}_{K - K_j \times K_j} \,
		\end{bmatrix}
	\end{equation}
	and Assumption \ref{Assumption_PT} requires that, for each $j = 1,\ldots, N_1$ and $t' > t_j^*$:
	
	$$\mathbb{E}[\eta_{jt'}] - \sum_{t=1}^{t_j^*} \frac{\mathbb{E}[\eta_{jt}]}{t_j^*} = \mathbb{E}[\eta_{it'}] -  \sum_{t=1}^{t_j^*} \frac{\mathbb{E}[\eta_{it}]}{t_{j}^*}, \forall i \in \{N_1+1,\ldots N_0\}\, ,$$
\end{example}

\begin{example}
	\label{example_pre_trends}
	Suppose we are interested in simunalteously conducting inference on pre- and post-treatment differential trends between treatment and control groups. We use the term ``post-treatment differential trends'' because, when there is possibly violation of parallel trends, the difference-in-differences estimand conflates treatment effects with differential trends in the untretated potential outcome. In this setting, with some abuse of notation, we define the building-block estimands as, for $j \leq N_1$ and $t=1\ldots T$:
	
	$$\alpha_{jt} = \mathbb{E}[y_{jt} - y_{jt^*_j}] - \frac{1}{N_1}\sum_{i=N_1+1}^{N}\mathbb{E}[y_{it}- y_{it^*_j}],$$ 
	i.e. the differential trend between unit $j$ and the average of controls with respect to the last pre-treatment period.\footnote{In this case, if Model \eqref{simple_did_equation} is valid, and Assumption \ref{Assumption_PT} holds, then this means that $\alpha_{jt}$ is the treatment effect for $t$ in the post-treatment periods, and it is zero for the pre-treatment periods. More generally, $\alpha_{jt}$ conflates treatment effects with violations of the parallel trends assumption.} Estimation of these building-block parameters may be conducted by setting
	
	$$A_j = \begin{bmatrix}
		\mathbb{I}_{t_j^*-1 } & -\boldsymbol{1}_{t_j^* -1 \times 1} & \boldsymbol{0}_{ t_j^*-1\times T-t_j^* } \\
		\boldsymbol{0}_{1 \times t_j^*-1} & 0 & \boldsymbol{0}_{1 \times T- t_j^*-1}\\
		\boldsymbol{0}_{T- t_j\times t_j^*-1} & -\boldsymbol{1}_{T-t_j^* \times 1} & \mathbb{I}_{T - t_j^*} 
	\end{bmatrix}\, .$$

We may then aggregate these parameters by ``length of exposure''. Specifically, we set $L_j = (1-t_j^*)$, $U_j = (T-t_j^*)$, $L = \min_{j \leq N_1} L_j$, $U= \min_{j \leq N_1} U_j$. For $k \in \mathbb{Z}$, we set $N_k =\sum_{j=1}^{N_1} (\mathbf{1}_{T - t_j^* \leq k} +  \mathbf{1}_{ (1-t_j^*)  \leq k})$. We then put:

$$B_j = \begin{bmatrix}
	\boldsymbol{0}_{(L_j-L)\times T} \\
	\operatorname{diag}(1/N_{L_j}, 1/N_{L_j+1}, \ldots 1/N_0, \ldots 1/N_{U_j}) \\
	\boldsymbol{0}_{(U - U_j)\times T} \, .
\end{bmatrix}$$

\end{example}

The next proposition shows our estimator is unbiased, albeit inconsistent. 

\begin{proposition}

Under Assumptions \ref{Assumption_sampling} and \ref{Assumption_PT}, the estimator $\widehat{\bar \alpha}$  is unbiased for $\bar \alpha$. Moreover,
 $$\widehat{\bar{\boldsymbol{\alpha}}} \rightarrow_p \bar{\boldsymbol{\alpha}} + \sum_{j=1}^{N_1} B_j A_j \boldsymbol{\eta}_j$$
when $N_0 \rightarrow \infty$ and $N_1 $ is fixed.

\end{proposition}

This proposition is equivalent to Proposition 1 from \cite{CT}. While the DID estimator is unbiased, it is not consistent, because the number of treated units is fixed. 

It would also be possible to use the not-yet-treated as part of the control group.  Since the asymptotic theory we consider in this paper considers $N_1$ fixed and $N_0 \rightarrow \infty$, this would not affect any of our asymptotic results. For the finite-sample result that the estimator   $\widehat{\bar \alpha}$  is unbiased,  we would have to adjust Assumption \ref{Assumption_PT} to include parallel trends for the not-yet-treated.

\section{Inference method}

The fact that $\widehat{\bar \alpha}$  is inconsistent poses some important challenges for inference. \cite{CT} propose an interesting alternative, in which the residuals from the control units can be used to estimate the distribution of the errors of the treated units. In their standard implementation, they consider a setting in which $(\boldsymbol{\eta}_1,...,\boldsymbol{\eta}_N)$  is iid.  \cite{FP} analyze the case in which  heteroskedasticity has a known structure that can be estimated from the data. For example, in case the treated units are state $\times$ time aggregates of individual level observations, then the idea is to estimate the heteroskedasticity using the residuals from the control units, and then use this estimated structure to make the residuals of the controls informative about the distribution of the errors of the treated.\footnote{\cite{CT} also consider in their appendix an alternative that allows for heteroskedasticity (based on a known variable) and spatial correlation (depending on an observed distance metric). The method proposed by \cite{FP} differs from this  alternative in that it does not require parametrization/estimation of the serial correlation structure, and that it does not assume normality.}

\cite{FP} focus on estimating a unidimensional parameter, and do not take into account  the recent advances in the analysis of staggered designs with heterogeneous effects. In their Section III.A, \cite{CT} discuss the possibility of extending their approach to settings in which there are heterogeneous treatment effects. In this note, we formalize a procedure to draw inference in this setting, taking into account that estimators for the building blocks $\alpha_{i,t}$ and $\alpha_{i,t'}$ are possibly correlated. We also show how parametric models for heteroskedasticity, as proposed by \cite{FP}, can be extended to this setting.  

\cite{CT} also suggest constructing confidence sets on multidimensional parameters in this setting by inverting a test statistic. However, confidence sets constructed from quadratic statistics (the typical choice, e.g. Wald test statistics) will be \textit{ellipsoidal}, which is difficult to both compute and visualize even for moderate dimensions of the target parameter \citep{Montiel2018}. This is particularly concerning for presenting confidence sets for event-study-like parameters. In contrast, we follow the recent literature on DID and event-studies in proposing uniform confidence bands for the target parameter \citep{Freyal2021, Pedro}. We show how such uniform confidence bands can be computed in a non-standard setting in which the estimator is not consistent. 


The next assumption imposes a parametric model for the heteroskedacity of $B_j A_j  \boldsymbol{\eta}_j$, and can be seen as an extension of the assumptions considered by \cite{FP}. Observe that Assumption \ref{Assumption_PT} implies that,  for each $j \leq N_1$,  $B_j A_j \boldsymbol{\eta}_i$ has equal mean for all $i \in \{j\} \cup \{N_1,\ldots, N\}$. Without loss, we assume that this mean is zero and that:

\begin{assumption}
	\label{Assumption_parametric_model}
	For each $j =1,\ldots N_1$, and for every $i \in \{j\} \cup \{N_1+1,\ldots, N\}$:
	
	$$B_j A_j \boldsymbol{\eta}_i = H_j(Z_i;\delta_j) \xi_{i,j},$$
	where $\delta_j \in \Delta_j \subseteq \mathbb{R}^{p_j}$ is an unknown parameter,  $H_j(\cdot;\cdot)$ is a known function such that $H_j(Z_i;\delta_j)$ is a positive definite $K \times K$ matrix for each $i \in \{j\} \cup \{N_1+1,\ldots, N\}$; and $\xi_{i,j} \overset{d}{=} \xi_{i',j}$ for every $i, i' \in \{j\} \cup \{N_1+1,\ldots, N\}$, where $\xi_{j,j}$ is a $K \times 1$ random variable  with a continuous distribution and $\mathbb{E}[\xi_{j,j}] = 0$.
\end{assumption}

 Note that, if we set $H_j(Z_i;\delta_j)$ as the identity matrix, then Assumption \ref{Assumption_parametric_model} would be implied by Assumption 2 from \cite{CT}, which states that $\boldsymbol{\eta}_i$ is iid. Therefore, all our results can be considered as extensions to the inference method proposed by \cite{CT} for the particular  case in which $H_j(Z_i;\delta_j)$ is the identity matrix.

\cite{FP} consider as a standard example the case in which outcomes $y_{it}$ come from aggregating data from $Z_i$ individuals in unit $i$ and time $t$. In this case, they show that, under a wide range of structures on the spatial and serial correlations within unit $i$,  the variance of $W_i = \frac{1}{T-t^\ast} \sum_{t=t^\ast+1}^T \eta_{it} - \frac{1}{t^\ast} \sum_{t=1}^{t^\ast} \eta_{it}$, as a function of $Z_i$, would have a parametric form given by $\mathbb{V}[W_i] = A + B \frac{1}{Z_i}$ for parameters $A ,B \geq 0$. In Appendix \ref{examples_models}, we show that, for this setting in which $y_{it}$ is the aggregate of  $Z_i$ individual-level observations, and under some homogeneity assumptions, this parametric form generalizes for our Example \ref{example_scalar} as $H_j(\cdot Z_i;\delta_j)^2=\mathbb{V}[B_j A_j \boldsymbol{\eta}_i] = \delta_{0j} + \delta_{1j} \frac{1}{Z_i}$, for $ \delta_{0j} , \delta_{1j} \geq 0$, and for our Examples \ref{example_es} and \ref{example_pre_trends} as $H_j(Z_i;\Lambda_{0j},\Lambda_{1j})^2 = \Lambda_{0j} +\frac{1}{Z_i}\Lambda_{1j}$ for $K \times K$ positive semidefinite matrices $\Lambda_{0j}$ and $\Lambda_{1j}$.

Assumption \ref{Assumption_parametric_model} suggests the following procedure for approximating the distribution of $\hat{\bar{\boldsymbol{\alpha}}} - \bar{\boldsymbol{\alpha}}$.

	\begin{enumerate}
		
		\item Estimate $\hat{\bar{\boldsymbol{\alpha}}}$. For each $j=1,\ldots, N_1$, compute and store $\widehat W_{i}(j) = B_j A_j \boldsymbol{Y}_i - \frac{1}{N_0}\sum_{i'=N_1+1}^{N}  B_j A_j \boldsymbol{Y}_{i'}  $ for every $i\in\{N_1+1,\ldots N\}$.
		
		\item Use the $\hat{W_{i}}(j)$ to construct estimators $\hat{\delta}_j$ of $\delta_{j}$, for $j=1,\ldots N_1$.
		
		\item For each $j =1,\ldots N_1$, compute the normalized residuals $\{\widetilde W_{i}(j)\}_{i=N_1+1}^N$, where $\widetilde W_{i}(j) = H_j(Z_i;\hat{\delta}_j)^{-1} \hat W_i(j)$.
		
		\item Do that $B$ times:
		
		\begin{enumerate}
			
			\item Draw a sequence of $N_1$ number with replacement from $\{N_1+1,...,N\}$, $(i_b(1),...,i_b(N_1))$.
			
			\item Compute and store $\hat{e}_b$, where  $$\hat{e}_b  = \sum_{j=1}^{N_1}  H_j(Z_j;\hat{\delta}_j) \widetilde{W}_{i_b(j)}(j)$$

		\end{enumerate}

	\end{enumerate}
	 
The next proposition provides conditions under which this approach correctly estimates the distribution of interest. In what follows, $\lVert \cdot \rVert$ denotes the spectral norm of a matrix, and $\lambda_{\operatorname{min}}(\cdot)$ is the smallest eigenvalue of a square matrix. We also denote by $F$ the distribution function of $\sum_{j=1}^{N_1} B_j A_j \boldsymbol{\eta}_j$.
\begin{proposition}
	\label{prop_distr}
	Suppose Assumptions \ref{Assumption_sampling}, \ref{Assumption_PT} and \ref{Assumption_parametric_model} hold. Suppose that, for each $j\leq N_1$, the estimators $\hat{\delta}_j$ of $\delta_j$ satisfy, as $N_0 \to \infty$, $\max_{i \in \{j\}\cup\{N_1+1,\ldots, N_0\}}\lVert H_j(Z_i;\hat\delta_{j}) - H_j(Z_i;\delta_{j}) \rVert \overset{p}{\to} 0$. Moreover, assume there exist $0 < \underline{h} \leq \overline{h} < \infty$ such that $\underline{h} \leq  \min_{ j\leq N_1, i \in \{N_1+1,\ldots N\}}\lambda_{\operatorname{min}}( H_j(Z_i;\delta_j)) $ and $\max_{ j\leq N_1, i \in \{j,N_1+1,\ldots N\}}\lVert H_j(Z_i;\delta_j)) \rVert \leq \overline{h} $, uniformly as $N_0 \to \infty$. We then have that $\hat{F}(c) \overset{p}{\to} F(c)$, uniformly over $c \in \mathbb{R}^K$, where $\hat{F}(c)$ is given by:
	
	$$\hat{F}(c) = \frac{1}{N_0^{N_1}} \sum_{i_1={N_1+1}}^{N} \sum_{i_2={N_1+1}}^{N} \ldots \sum_{i_{N_1}={N_1+1}}^{N} \mathbf{1}\left\{\sum_{j=1}^{N_1}H(Z_{j};\hat{\delta}_j) \widetilde{W}_{i_j}(j)\leq c\right\}\, .$$
	\begin{proof}
		See Appendix \ref{proof_prop}.
	\end{proof}
\end{proposition}
\begin{remark}
{Assumption \ref{Assumption_parametric_model} requires the matrices $H_j(Z_i;\delta_j)$ to be positive definite, whereas in Examples \ref{example_es} and  \ref{example_pre_trends}, some of the entries of $B_jA_j \boldsymbol{\eta}_i$ are degenerate. Our results immediately follow in these settings by modifying Assumption \ref{Assumption_parametric_model} and the smallest eigenvalue assumption in Proposition \ref{prop_distr} to hold for the submatrix of $H_j(Z_i;\delta_j)$ which stores the covariance matrix of the nondegenerate part of $B_jA_j \boldsymbol{\eta}_i$. We do not state the assumptions this way for ease of exposition.}
\end{remark}

\begin{remark}
	Propososition \ref{prop_distr} requires estimators $\hat \delta_j$ of the parametric model. These may be obtained by solving:
	
	$$\min_{d_j\in \Delta_j } \sum_{s=N_1+1}^{N}\lVert\hat{W}_s(j)\hat{W}_s(j)' - H_j(Z_i;d_j)H_j(Z_i;d_j)' \rVert_F^2,$$
	where $\lVert \cdot \rVert_F$ denotes the Froebenius norm. The resulting estimator satisfies the assumption required in  Proposition \ref{prop_distr} under standard conditions. Indeed, in our leading example where the parametric model comes from aggregation from individual-level data, uniform consistency of the estimated variances is ensured if $\lVert \hat{\Lambda}_{0,j} - {\Lambda}_{0,j} \rVert \overset{p}{\to}0$ and $\lVert \hat{\Lambda}_{1,j} - {\Lambda}_{1,j} \rVert \overset{p}{\to}0$ for all $j \leq N_1$, which can be ensured by relying on standard arguments on least-squares estimation.
\end{remark}
\begin{remark}
	We also note that, for our proposed algorithm to work, we require that the \textbf{estimated} $\hat{H}_j(Z_i, \hat{\delta}_j)$, $i > N_1$, be invertible. This occurs with probability approaching one in our setting. However, to ensure invertibility in finite samples, it may be convenient to include a penalisation on small values of the smallest singular value of $\hat{H}_j(Z_i, \hat{\delta}_j)$. Provided the penalty vanishes asymptotically, the estimator retains the required properties by Proposition \ref{prop_distr}.
\end{remark}

Proposition \ref{prop_distr} provides conditions under which the proposed algorithm recovers the distribution of the test statistic. This can be used to construct confidence intervals for the target parameters. If the parameter of interest is unidimensional ($K=1$), an asymptotically valid  $(1-\alpha)$ confidence intervals for $\bar{\alpha}$ can be constructed as:

$$[ \hat{\bar{\alpha}}  - \hat q_{1-\alpha}, \hat{\bar{\alpha}} + \hat q_{1-\alpha}]\, ,$$
where $\hat{q}_{1-\alpha}$ is the $1-\alpha$ empirical quantile of $|\hat{e}_b|$. Alternatively, if $K>1$,$(1-\alpha)$ uniform confidence bands may be constructed as:

$$\prod_{s=1}^K[\boldsymbol{\hat{\bar\alpha}}_s - \hat \iota_s \hat q_{1-\alpha}, \boldsymbol{\hat{\bar\alpha}}_s + \hat \iota_s q_{1-\alpha}],$$
where $\hat \iota_s$ are normalizing constants, and $\hat{q}_{1-\alpha}$ is the $1-\alpha$ empirical quantile of $ \max_{s=1,\ldots, K} |\hat{e}_{b,s}/\hat \iota_{s}|$. Typical choices include $\hat \iota_{s}=1$ for all $s$, which leads to a constant-width uniform band; and $\hat \iota_{s} = \hat \sigma_s$, where  $\hat \sigma_s^2$ is the estimator of the variance of $\hat{\bar{\alpha}}_s$ using the simulated draws. 

We summarise the  discussion in the corollary below.

\begin{corollary}
	\label{cor_confidence}
	Suppose Assumptions \ref{Assumption_sampling}, \ref{Assumption_PT}, \ref{Assumption_parametric_model} and the conditions in Proposition \ref{prop_distr} hold. Let $\mathcal{C}$ be a confidence interval constructed as:
	$$\mathcal{C} = \prod_{s=1}^K[\boldsymbol{\hat{\bar\alpha}}_s - \hat \iota_s \hat q_{1-\alpha}, \boldsymbol{\hat{\bar\alpha}}_s + \hat \iota_s q_{1-\alpha}],$$
	where the $\iota_s$ are such that $\hat\iota_s \overset{p}{\to} \iota_s > 0$ as $B,N_0\to \infty$. We then have that:
	
	$$\lim_{N_0,B\to \infty} \mathbb{P}[\boldsymbol{{\bar{\alpha}}} \in \mathcal{C}] = 1- \alpha.$$
	
	Specifically, if $\hat\iota_s = \hat{\sigma_s}$, then $\hat{\sigma}_s^2 \overset{p}{\to} \sigma_s^2$ as $B,N_0\to \infty$.
	
	\begin{proof}
		See Appendix \ref{proof_cor}.
	\end{proof}
\end{corollary}



\singlespace

\renewcommand{\refname}{References} 

\bibliographystyle{apalike}
\bibliography{bib/bib.bib}

\appendix
\section{Parametric models for examples}
\label{examples_models}
\subsection{Panel data}
Suppose the $y_{it}$ are obtained from aggregating individual-level data from $Z_i$ individuals. In this case, we may write:

$$\boldsymbol{\eta_i} = \boldsymbol{f}_i + \frac{1}{Z_i} \sum_{s=1}^{Z_i} \boldsymbol{\epsilon}_{s,i},$$
where $\boldsymbol{f}_i$ are group shocks, $\boldsymbol{\epsilon}_{s,i}$ are idiosyncratic shocks, and, without loss of generality, $\operatorname{cov}(\boldsymbol{f}_i, \boldsymbol{\epsilon}_{s,i}) = 0$. If we assume $\boldsymbol{\eta}_i$ to be iid across $i$ and $\boldsymbol{\eta}_{s,i}$ to be iid across $s$ and $i$, we have that:

$$\mathbb{V}[\boldsymbol{\eta}_i] = V_0 + \frac{V_1}{Z_i}$$

For positive semidefinite matrices $V_0$ and $V_1$. For given choices of $A_j$ and $B_j$, we thus obtain the model:

$$H_j(Z_i;\delta_j)^2 = \Lambda_{0j} + \frac{\Lambda_{1j}}{Z_i},$$
where $\Lambda_{0j}$ and $\Lambda_{1j}$ are semidefinite $K \times K$ matrices.

\subsection{Repeated cross-sections}
The model in the previous section is especially suited for panel data, since it allows for serial correlation in idiosyncratic shocks. In the case where the data $y_{jt}$ is constructed from repeated cross-sections with $Z_{jt}$ individuals, a more parsimonious model can be obtained. Indeed, in this case, we consider the model:

$$\boldsymbol{\eta}_i = \boldsymbol{f}_i + \begin{bmatrix}
	\frac{1}{Z_1}\sum_{s=1}^{Z_1} \epsilon_{s,i,1}
	\\
	\vdots \\\
	\frac{1}{Z_T}\sum_{s=1}^{Z_T} \epsilon_{s,i,T}
\end{bmatrix}\, .$$
where we assume that $\boldsymbol{f}_i$ is iid across $i$; for each $t$, $\epsilon_{s,i,t}$ is iid across $s$ and $i$; and that $\epsilon_{s,i,t}$ $\epsilon_{s',i',t'}$ are independent for $t \neq t'$. This leads to:

$$\mathbb{V}[\boldsymbol{\eta}_i]= V_0 + \operatorname{diag}(\sigma^2_1/Z_{i1},  \sigma^2_2/Z_{i2}, \ldots, \sigma^2_T/Z_{iT}),$$

In Example \ref{example_es}, this model leads to the following parametrization:

$$H_j(Z_j;\delta_j)^2 = \Lambda_{0j} + \frac{\omega_{0j}}{Z_{i,t_j^*}} \begin{bmatrix}
	\boldsymbol{1}_{K_j\times K_j} & \boldsymbol{0}_{K_j \times K - K_j} \\
	\boldsymbol{0}_{K - K_j \times K_j} & \boldsymbol{0}_{K - K_j \times K - K_j}
\end{bmatrix}+ \operatorname{diag}\left(\frac{\omega_{1j}}{Z_{i,t_j^*+1}}, \frac{\omega_{2j}}{Z_{i,t_j^*+2}}, \ldots, \frac{\omega_{K_j,j}}{Z_{i, K_j}}, 0, 0, \ldots, 0\right)$$
\section{Proofs of main results}
\subsection{Proof of Proposition \ref{prop_distr}}
\label{proof_prop}
First, we note that, by Weyl's inequality and the eigenvalue and spectral norm assumptions in the statement of the theorem, $\mathbb{P}[\cap_{j=1}^{N_1} \cap_{i=N_1+1}^{N}\{H_j(Z_i;\hat \delta_j)^{-1} \text{ exists}\}] \to 1$. Without loss, we assume these matrices to be always invertible.\footnote{Otherwise, premultiply random variables that depend on an inverse by an indicator of the event where all matrices are invertible. The following argument remains essentially unchanged, except for additional $o_p(1)$ terms appearing in the derivations. }
Let:

$$\tilde{\boldsymbol{e}} = \sum_{j=1}^{N_1} H_j(Z_j;\hat \delta_j) \tilde W_{i^*_j}(j),$$
where the $i^*_j$ are iid draws from $\operatorname{Uniform}(\{N_1+1,\ldots, N\})$, independently from $\{\boldsymbol{\eta}_i\}_{i}$. Do further define 

$${\boldsymbol{e}} = \sum_{j=1}^{N_1} H_j(Z_j; \delta_j) \xi_{i^*_j,j}.$$

We show  $\lVert\hat{\boldsymbol{e}} - \boldsymbol{e}\rVert = o_p(1)$. Note that:

\begin{equation}
	\label{eq_inequality}
	\begin{aligned}
\lVert\hat{\boldsymbol{e}} - \boldsymbol{e}\rVert \leq \sum_{j=1}^{N_1} \lVert H_j(Z_j;\hat{\delta}_j) - H_j(Z_j;{\delta}_j)\rVert \lVert \tilde{W}_{i^*_j}(j) - {\xi}_{i^*_j,j} \rVert + \sum_{j=1}^{N_1} \lVert H_j(Z_j;\hat{\delta}_j) - H_j(Z_j;{\delta}_j)\rVert \lVert {\xi}_{i^*_j,j} \rVert + \\
\sum_{j=1}^{N_1} \lVert H_j(Z_j;{\delta}_j)\rVert \lVert H_j(Z_j;\hat{\delta}_j) - H_j(Z_j;{\delta}_j)\rVert
\end{aligned}
\end{equation}

Fix $j \leq N_1$. First, by the Assumption in the statement of the Proposition, $\lVert H_j(Z_j;\hat{\delta}_j) - H_j(Z_j;{\delta}_j)\rVert =o_p(1)$. Moreover, by Assumption \ref{Assumption_parametric_model}, $\xi_{i_j^*,j} \overset{d}{=} \xi_{j,j}$, implying $\lVert\xi_{i_j^*,j}\rVert = O_p(1)$. Finally, letting $\mathbb{E}_*$ denote expectations with respect to the distribution of $(i_1^*,\ldots, i_{N_1}^*)$, with $\{\boldsymbol{\eta}_i\}$ fixed, we observe that:

\begin{equation}
	\begin{aligned}
	\mathbb{E}_* \lVert \tilde{W}_{i^*_j}(j) - {\xi}_{i^*_j,j} \rVert  = \frac{1}{N_0}\sum_{i=N_1+1}^{N} \lVert \tilde{W}_{i}(j) - {\xi}_{i,j} \rVert	 \leq\\\max_{i \in \{N_1+1,\ldots N\}} \lVert H_j(Z_i;\delta_j)^{-1} - H_j(Z_i;\hat \delta_j)^{-1} \rVert\left(\left\lVert \frac{1}{N_0} \sum_{i=N_1+1}^{N} B_jA_j\boldsymbol{\eta}_i\right \rVert+	\frac{1}{N_0}\sum_{i=N_1+1}^{N} \left\lVert B_j A_j \boldsymbol{\eta}_i\right\rVert\right) + 
	\\
	\max_{i \in \{N_1+1,\ldots N\}} \lVert H_j(Z_i;\delta_j)^{-1} \rVert \left\lVert \frac{1}{N_0} \sum_{i=N_1+1}^{N} B_jA_j\boldsymbol{\eta}_i\right \rVert = o_p(1) 
	\end{aligned}
\end{equation}
where we use the law of large numbers, the fact that, for a symmetric invertible matrix $A$, $\lVert A^{-1} \rVert = 1/\lambda_{\operatorname{min}}(A)^2$; and that for invertible matrices $A$ and $B$, 

$$\lVert A^{-1} - B^{-1} \rVert \leq \left(\frac{1}{\lVert B \rVert} \right)\frac{\lVert A - B \rVert}{\lVert B \rVert  - \lVert A - B \rVert}\, .$$

By the conditional Markov inequality, $\mathbb{P}_*[\lVert \tilde{W}_{i^*_j}(j) - {\xi}_{i^*_j,j} \rVert>\epsilon ] = o_p(1)$ for every $\epsilon > 0$. Iterated expectations and the bounded convergence theorem imply $\lVert \tilde{W}_{i^*_j}(j) - {\xi}_{i^*_j,j} \rVert = o_p(1)$. We have thus shown that, for fixed $j \leq N_1$, the sum of terms on the left-hand side of \eqref{eq_inequality} associated with such $j$ is $o_p(1)$. Since the chosen $j$ was arbitrary, we conclude the whole term is $o_p(1)$, which proves that $\lVert\tilde{\boldsymbol{e}} - \boldsymbol{e}\rVert = o_p(1)$. 

Next, define $$\tilde{\boldsymbol{g}} = \sum_{j=1}^{N_1} H_j(Z_j;\hat \delta_j) \tilde W_{s^*_j}(j),$$
where $s^*_j$ are iid draws from $\operatorname{Uniform}(N_1+1,\ldots, N)$, independently from both the $i^*_j$ and the $\boldsymbol{\eta}_i$. Similarly, let:

$$\boldsymbol{g} = \sum_{j=1}^{N_1}  \xi_{s^*_j,j}.$$
 
Note that $\lVert\tilde{\boldsymbol{g}} - \boldsymbol{g}\rVert = o_p(1)$. Next, we claim that $(\boldsymbol{e}',\boldsymbol{g}')' \overset{d}{\to} F \otimes F$. To see this, observe that the event $E = \{\exists j \neq j': i_j^* = i_{j'}^* \text{ or } s_j^* = s_{j'}^* \} \cup \{\exists j, j': i_j^* = s_{j'}^*\}$ is such that $\mathbb{P}[E]  \to 0 $. We thus have that, for any $c_1, c_2 \in \mathbb{R}^K$:
 
 \begin{equation*}
 	\begin{aligned}
 		\mathbb{P}[\boldsymbol{e}\leq c_1, \boldsymbol{g} \leq c_2 ] = F(c)^2 \mathbb{P}[E^c] + \mathbb{P}[\{\boldsymbol{e} \leq c_1 \} \cap \{\boldsymbol{g} \leq c_2 \} \cap E]  =  F(c)^2 + o(1)		\, .
 	\end{aligned}
 \end{equation*} 

Application of Slutsky lemma then yields $(\tilde{\boldsymbol{e}}',\tilde{\boldsymbol{g}}')' \overset{d}{\to} F\otimes F$. It then follows from Theorem 15.2.3 of \cite{Lehmann2005} that $\hat{F}(c) \overset{p}{\to} F(c)$ for every $c \in \mathbb{R}^K$. Uniform convergence follows from the fact that $F$ is continuous \citep[p. 339]{Vaart1998}.

\subsection{Proof of Corollary \ref{cor_confidence}}
\label{proof_cor}
Let $\hat{F}_B$ denote the empirical distribution obtained from running the proposed algorithm $B$ times. Denoting by $\mathbb{E}_*$ the expectation with respect to the randomness of the algorithm and with the data fixed, we have that, for each $c \in \mathbb{R}^K$:

$$\mathbb{E}_*[\hat{F}_B(c)] = \hat{F}(c)\, ,$$

$$\mathbb{V}_*[\hat{F}_B(c)] = \frac{\hat{F}(c)(1-\hat{F}(c))}{B}\, .$$

It then follows from Proposition \ref{prop_distr} that, for each $\epsilon > 0$:

$$\lim_{N_0, B \to \infty}\mathbb{P}_*[|\hat{F}_B(c) - F(c)|>\epsilon] = 0.$$

The bounded convergence theorem and  the fact that $F$ is continuous thus imply that:

$$\sup_{c \in \mathbb{R}^K}|\hat{F}_B(c) - F(c) |= o_p(1)$$

Let $\hat \iota = \operatorname{diag}(\hat \iota_1, \ldots \hat \iota_s)$. Since $F$ is continuous and $\hat \iota \overset{p}{\to} \iota$, we have that:

$$\sup_{c \in \mathbb{R}^K}|\hat F_B(\hat \iota c) - F(\iota c)| = o_p(1),$$
and it follows from the continuous mapping theorem that:

$$\sup_{c \in \mathbb{R}}|\hat G_B(c) - G(c)| = o_p(1)\,,$$
where $\hat G_B$ is the empirical distribution of $\max_{s=1,\ldots,K} |\hat e_{s,b}|/|\hat \iota_s|$, and $G(c) = \int_{\boldsymbol{[-c,c]^K}} F(\iota dx) $. By Lemma 21.2 of \cite{Vaart1998}, it follows that $\hat{q}_{1-\alpha}$ converges in proability to the $(1-\alpha)$ quantile of $G$. Sjnce by Slutsky theorem ${\hat \iota}^{-1} (\boldsymbol{\hat{\bar{\alpha}}-\boldsymbol{\bar{\alpha}}})$ converges weakly to $F(\iota c)$, we have, by the continuous mapping theorem:

\begin{equation}
	\begin{aligned}
		\mathbb{P}[\boldsymbol{{\bar{\alpha}}} \in \mathcal{C}] = \mathbb{P}[  \mathbf{1}\hat{q}_{1-\alpha}\leq {\hat \iota}^{-1} (\boldsymbol{\hat{\bar{\alpha}}} -\boldsymbol{\bar{\alpha}}) \leq  \mathbf{1}\hat{q}_{1-\alpha}] = \\
		 \mathbb{P}[{\hat \iota}^{-1} (\boldsymbol{\hat{\bar{\alpha}}} -\boldsymbol{\bar{\alpha}}) \leq \mathbf{1}\hat{q}_{1-\alpha}] - \sum_{\boldsymbol{\kappa} \in \{-1,1\}^K: \exists \boldsymbol{\kappa}_s = 1} \mathbb{P}[\operatorname{diag}(\boldsymbol{\kappa}){\hat \iota}^{-1} (\boldsymbol{\hat{\bar{\alpha}}} -\boldsymbol{\bar{\alpha}}) \leq -\operatorname{diag}(\boldsymbol{\kappa}) \hat{q}_{1-\alpha} ]\to \\ \int_{[-q_{1-\alpha}\boldsymbol{1}, q_{1-\alpha}\boldsymbol{1}]} F(\iota dx) = 1-\alpha\,.
	\end{aligned}
\end{equation}
The last assertion of the corollary is immediate and therefore not proved.

\end{document}